\documentclass[english,aps,preprint]{revtex4}%
\usepackage[T1]{fontenc}
\usepackage[latin9]{inputenc}
\usepackage{amsmath}
\usepackage{amssymb}
\usepackage{babel}
\usepackage{amsfonts}
\usepackage{graphicx}%

\begin{document}
\title{Space-time resolved quantum field approach to Klein tunneling dynamics across
a finite barrier}
\author{M.\ Alkhateeb}
\author{A. Matzkin}
\affiliation{Laboratoire de Physique Théorique et Modélisation, CNRS Unité 8089, CY Cergy
Paris Université, 95302 Cergy-Pontoise cedex, France}

\begin{abstract}

We investigate Klein tunneling through finite potential barriers with space-time resolved solutions to
relativistic quantum field equations. We find that no particle actually tunnels through a finite supercritical barrier, even in the case of resonant tunneling. The transmission is instead mediated by modulations in pair production rates, at each edge of the barrier, caused by the incoming electron. We further examine the effect of the barrier?s width on the numbers of produced pairs in the fermionic case (characterized by saturation) and in the bosonic case (characterized by exponential superradiance). This work paves the way to precise studies of the radiating dynamics of supercritical barriers, and could be applied to certain analogs of Klein tunneling observed in systems modeled by relativistic wave equations.
\end{abstract}
\maketitle


Vacuum pair production in a strong supercritical external field is one of the
basic elementary processes predicted by relativistic quantum theory
\cite{ruffiniR,reviewR}. Understanding its precise dynamics is important not only for
fundamental reasons, but also due to current efforts aiming to experimentally
observe pair production in the intense laser facilities currently under
construction \cite{dunneR,hu}. To this end, several works employing different
approaches have examined field configurations that would optimize pair
production \cite{shabaev,gies,grobe-opt,lv}.

A related aspect concerns the interaction of a charged particle with the pair
production process as the particle scatters on an inhomogenous supercritical
field. This is well-known to give rise to ``Klein
tunneling'' \cite{calogeracos}, that in a first quantized
framework appears as undamped propagation inside the potential region. The
precise dynamics of this interaction has remained controversial even in the
case of a simple electrostatic step, a situation giving rise to the so-called
Klein paradox \cite{greiner}. Time-independent first quantized approaches are
generally misleading \cite{alkhateeb2021}, but even stationary quantum field
theory (QFT) methods failed to reach consensus (e.g. the choice of asymptotic
``in''  and ``out'' fields \cite{nikishov,hansen,gavrilov,kleinert}). A
numerical space-time resolved QFT approach \cite{grobe2004} was instrumental in
computing the precise dynamics of the interaction between the incoming particle
and the pair production for a step, leading to a reinterpretation of the
fermionic Klein paradox in terms of Pauli blockade of vacuum pair production.

For a particle impinging on a supercritical electrostatic barrier of finite
width, the computation of the time-dependent pair creation rates and of the
dynamics inside the barrier is expected to be more involved; in particular the
asymptotic field operators only contain particles, which has lead some authors
\cite{calogeracos,mckellar} to conjecture that a symmetric supercritical
barrier once formed cannot radiate. While genuine Klein tunneling has yet to
be observed for elementary particles, the physics encapsulated in the first
quantized Dirac equation has been used as an effective model in other areas,
leading to the experimental observation of Klein tunneling in graphene
heterojunctions \cite{cmo}, with photonic crystals \cite{kgs}, trapped ions \cite{bco} or cold quantum gases
\cite{oo}.

In this work, we implement a time-dependent space-resolved QFT treatment in
order to compute the detailed dynamics of Klein tunneling for fermions and
bosons across a finite barrier. In the fermionic case, the calculations will lead us to propose a mechanism
accounting for the undamped transmission characterizing Klein tunneling: due to the exchange symmetry, pair production appears only as a transient effect when the field is turned on. A particle incoming on the saturated barrier then induces modulations in the anti-particle density which in turn triggers production of the transmitted particle. Hence no particle actually tunnels inside the barrier, even in the
resonant case of nearly full transmission. In the bosonic case, a barrier amplifies pair production, with the
anti-boson charge oscillating inside the barrier increasing exponentially
each time it scatters on an edge.

\begin{figure}[b]
	\centering \includegraphics[width=8.4cm]{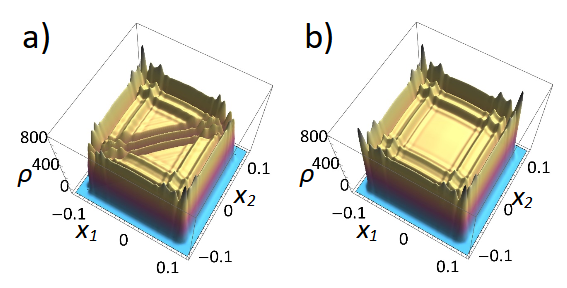} \label{fig1}%
	\caption{(a) The spatial density $\rho(x_1,x_2)$ of positron couples at positions $x_1$ and $x_2$ created inside the barrier
		at time $2 \times 10^{-3}$ expressed in atomic units (a.u., defined by $\hbar=m=1$,$c=137.036$, where $m$ is the electron mass); the barrier has width $L=0.2$ a.u., height $V_0=3mc^2$ and the smoothness has been set to $\epsilon=0.3/c$. (b) Same as (a) but without  the exchange interaction terms in the computations, see Eq. (\ref{2p1}).}%
\end{figure}

The barrier is modeled as a one-dimensional background external field with
negligible backreaction \cite{gitman-back}. The fermionic pair production rate
is obtained from field operators $\Psi$ expanded as%
\begin{equation}
\Psi(x,t)=\int dp\left(  b_{p}(t)\phi_{p}(x)+d_{p}^{\dagger}(t)\varphi
_{p}(x)\right)  . \label{field1}%
\end{equation}
$\phi_{p}$ and $\varphi_{p}$ are resp.\ the positive and negative energy
spinor eigenfunctions of the field-free Dirac Hamiltonian $H_{0}=-i\hbar
c\alpha_{x}\partial_{x}+\beta mc^{2}$ with eigenvalues $\pm\left\vert
E_{p}\right\vert =\pm\sqrt{p^{2}c^{2}+m^{2}c^{4}}$ ($\alpha$ and $\beta$ are
the usual Dirac matrices
\footnote{We will consider the usual one effective spatial dimension
approximation, neglecting spin-flip and replacing $\alpha_{x}$ and $\beta$ by
the Pauli matrices $\sigma_{1}$ and $\sigma_{3}$ respectively.}, $m$ the
electron mass and $c$ the light velocity) $b_{p}(t)$ (resp. $d_{p}(t))$ is the
annihilation operator for a particle (resp.\ antiparticle), and $b_{p}%
^{\dagger}(t)$ and $d_{p}^{\dagger}(t)$ are the corresponding creation
operators obeying the usual commutation relations, i.e. the only non-zero
equal time anti-commutators are $[b_{p},b_{k}^{\dagger}]_{+}=[d_{p}%
,d_{k}^{\dagger}]_{+}=\delta(p-k).$ The time-dependence of the creation and
annihilation operators is obtained by noting that $\Psi$ obeys the Dirac
equation with the full Hamiltonian $H=H_{0}+V(x)$ where $V(x)$ is the
background potential. It then follows that \cite{qft-rev}%

\begin{align}
b_{p}(t)  &  =\int dk\left(  U_{\phi_{p}\phi_{k}}(t)b_{k}(0)+U_{\phi
_{p}\varphi_{k}}(t)d_{k}^{\dagger}(0)\right) \label{bogu1}\\
d_{p}^{\dagger}(t)  &  =\int dk\left(  U_{\varphi_{p}\phi_{k}}(t)b_{k}%
(0)+U_{\varphi_{p}\varphi_{k}}(t)d_{k}^{\dagger}(0)\right)  . \label{bogu2}%
\end{align}

The time-evolved amplitudes, defined by 
\begin{equation}
U_{\phi_{k}\varphi_{p}}%
(t)\equiv\left\langle \phi_{k}\right\vert \exp\left(  -iHt/\hbar\right)
\left\vert \varphi_{p}\right\rangle 	
\end{equation}
 are computed numerically on a
discretized space-time grid by relying on a split operator \cite{split} method: the evolution operator is split into a kinetic part propagated in momentum space and a potential-dependent part solved in position space \cite{SP}.

Let us first consider vacuum pair production in a quasi rectangular potential
barrier of width $L;$ for definiteness we take $V(x)=\frac{V_{0}}{2}\left[
\tanh(\left(  x+L/2\right)  /\epsilon)-\tanh(\left(  x-L/2\right)  /\epsilon)
\right]  \ $where $V_{0}>2mc^{2}$ is supercritical and $\epsilon$ is a
smothness parameter. The electron density is obtained from the vacuum
expectation value $\rho_{el}(x,t)=\left\langle 0\right\vert \Psi_{el}%
^{\dagger}(x,t)\Psi_{el}(x,t)\left\vert 0\right\rangle $, where $\Psi_{el}$ is
the positive energy part of Eq. (\ref{field1}). The positron density
$\rho_{pos}(x_{1},x_{2},t)$ is obtained similarly from $\Psi_{pos}$ defined as
the positive energy part of the field conjugate to the one given by Eq.
(\ref{field1}) \cite{schweber}. It should be stressed that such densities (in particular
the positron densities) represent particle densities in a field-free basis \cite{add-r}. This 
corresponds to an experiment in which the particles would be counted after switching off the
field instantaneously.

Since a potential of the form $V(x)$ typically creates
two\ pairs simultaneously (one at edge of the barrier, where the field is
highly inhomogeneous), it is particularly instructive to compute the
2-particle density for a joint electron or positron creation at $x_{1}$ and
$x_{2}$,%
\begin{equation}
\rho_{\text{a}}(x_{1},x_{2},t)=\left\langle 0\right\vert \Psi_{\text{a}%
}^{\dagger}(x_{1},t)\Psi_{\text{a}}^{\dagger}(x_{2},t)\Psi_{\text{a}}%
(x_{2},t)\Psi_{\text{a}}(x_{1},t)\left\vert 0\right\rangle /2! \label{2el}%
\end{equation}
where 'a' stands for $pos$ or $el$. Note that by construction $\rho
(x_{1},x_{2})$ also counts the number of multiple pairs created on the same
edge of the barrier, although the probability for multiple pair creations at a
single field inhomogeneity is expected to decrease exponentially
as the multiplicity number increases \cite{oldmulti,grobe-multi} (but see \cite{lv-bauke}).
\begin{figure}[tb]
	\centering \includegraphics[width=8.8cm]{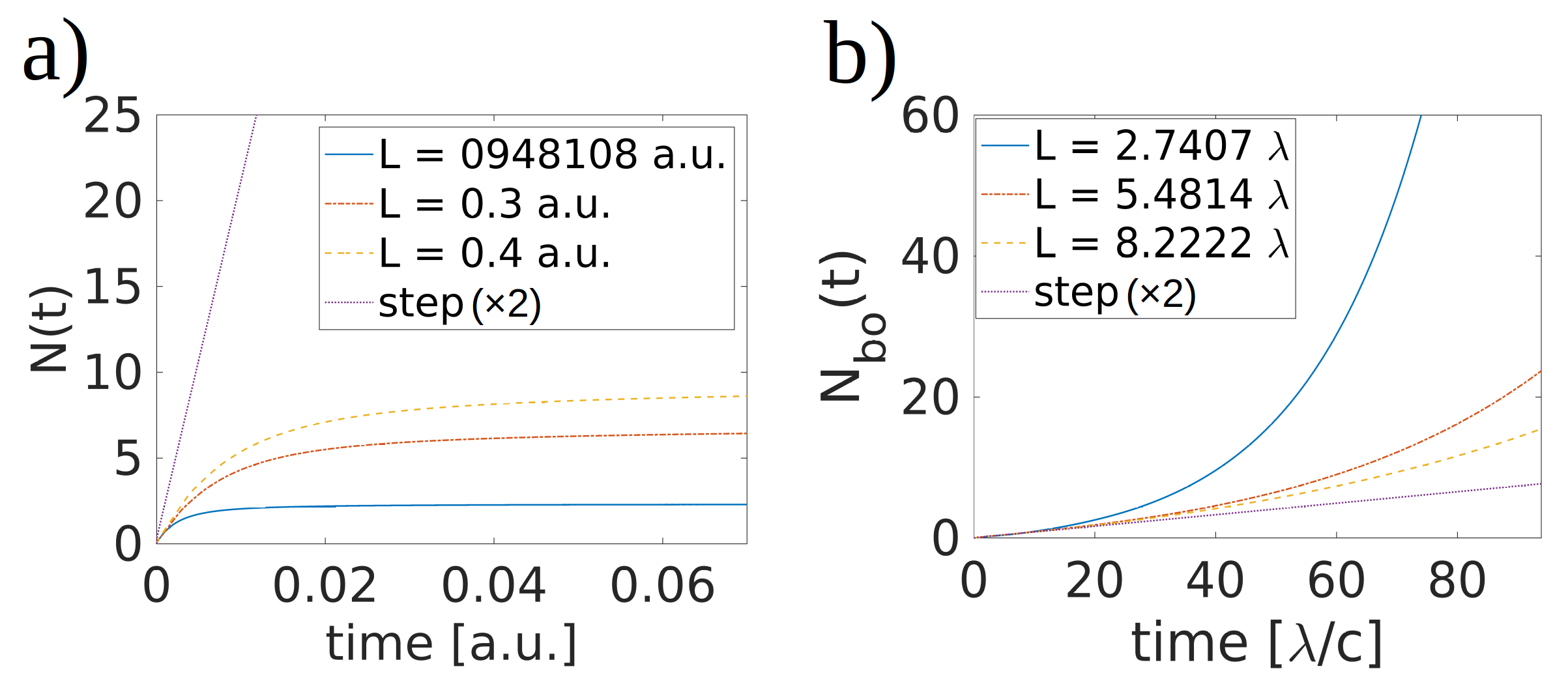} \label{fig1a}%
	\caption{ (a) Time-dependence (in a. u.) of the total number of created electron-positron pairs $N(t)$ for a fermionic barrier  with the parameters given in Fig. 1 except for the width $L$, indicated in the legend. The rate vanishes except in the limit of a potential step (note that in order to compare the step and finite barrier cases, the curve shown for a step has been multiplied by two). (b) The total number of created boson-anti-boson pairs $N_{bo}(t)$ for a massive bosonic field obeying the Klein-Gordon equation interacting with a background potential barrier  with $V_0=3m_{bo} c^2$, $\epsilon=0.3/c$, and varying widths; $m_{bo}$ is the boson mass which renormalizes the atomic units, so that $L$ and $t$ are given in units of $\lambda=m_{bo}\hbar/c$ and $\lambda/c$ respectively.}%
\end{figure}
The two-particle density (\ref{2el}) can be shown to be written as \cite{SP}%
\begin{equation}
\rho_{\text{a}}(x_{1},x_{2},t)=\rho_{\text{a}}(x_{1},t)\rho_{\text{a}}%
(x_{2},t)/2-\rho_{\text{a}}^{int}(x_{1},x_{2,}t), \label{2p1}%
\end{equation}
indicating that the number of created pairs is affected by the exchange
interaction terms encapsulated in $\rho^{int}$. This is portrayed in
Fig.\ 1 for the positron density inside the barrier. As the positrons
produced at each edge propagate towards the opposite side, $\rho_{pos}^{int}$
becomes of the same order of magnitude as the factorized density
$\rho_{pos}(x_1)\rho_{pos}(x_2)$, so that the exchange interaction significantly reduces the
number of couples inside the barrier.

The number $N(t)$ of created
positron-electron pairs, obtained by integrating either $\rho_{pos}(x,t)$ or $\rho_{el}(x,t)$ over all space (i. e. essentially inside the barrier for $\rho_{pos}$, outside for $\rho_{el}$)
is shown in Fig.\ 2(a) for different barrier widths. $N(t)$ is seen to
saturate after a transient time that depends on the barrier width: at short
times the number of produced pairs increases similarly to a step potential,
but a fermionic symmetric barrier will not radiate after this transient
period, as expected from the asymptotic behavior of the evolution amplitudes
$U_{\phi_{k}\varphi_{p}}(t)$ \cite{prep}.

We can now examine how an electron colliding on the supercritical barrier
interacts with the fermionic pair production process. We will consider
resonant Klein tunneling, that has been widely investigated in the condensed
matter analog of quasi-particles undergoing nearly full transmission through
electro-static barriers in graphene \cite{graphene}. The field operators are
again given by Eqs. (\ref{field1})-(\ref{bogu2}), but the spatial densities
 are obtained from the expectation values
 \begin{equation}
\rho_{\text{a}}(x)= 	\left\langle \zeta\right\vert
 	\Psi_{\text{a}}^{\dagger}(x)\Psi_{\text{a}}(x)\left\vert \zeta\right\rangle 
 	\label{densy}
 \end{equation}  
where $\left\vert\zeta\right\rangle =\int dp\zeta_{p}(p_{0},x_{0})b_{p}^{\dagger}(0)\left\vert
0\right\rangle $ represents the initial electron wavepacket centered at
$x_{0}$ with mean momentum $p_{0}$; for definiteness we take $\zeta_{p}%
(p_{0},x_{0})\propto e^{-\Delta^{2}(p-p_{0})^{2}-ipx_{0}/\hbar}$ corresponding
to a Gaussian wavepacket of width $\Delta$, similarly to previous works
dealing with the Klein paradox due to a potential step
\cite{grobe2004,keitel2009}. The resonant barrier condition \cite{calogeracos2}
\begin{equation}
	\left( p_{0}^{2}+V^{2}/c^{2}-2V\sqrt{p_{0}^{2}/c^{2}+m^{2}}\right) ^{1/2}=\frac{k\pi }{2L}
	\label{res-c}
\end{equation}
(which holds for a rectangular barrier, corresponding to the potential slope parameter $\epsilon
\rightarrow0$) is obtained by maximizing the transmitted Dirac current
relative to the incident one ($k$ is an integer).\ In the first quantized
approach, this means that the transmission amplitude $T(p_{0})$ for stationary
solutions of the Dirac equation is unity, so that if the wavepacket is
sufficiently narrow in momentum, it will nearly entirely tunnel through the
barrier \cite{alkhateeb2021}.

However in the present more fundamental second quantized framework, the incoming
electron interplays with the particles created by the supercritical potential.
The electronic spatial density given by Eq. (\ref{densy}) can be shown  \cite{SP} to take the form, by using Eqs. (\ref{field1})-(\ref{bogu2}), $\rho_{el}%
=\left\langle 0\right\vert \Psi_{el}^{\dagger}\Psi
_{el}\left\vert 0\right\rangle +\rho_{el}^{\zeta}$ where
\begin{equation}
\rho _{el}^{\zeta }(x)=\left\vert\int dpdk \zeta (p)U_{\phi _{k}\phi _{p}}(t)\phi _{k}(x)\right\vert ^{2}
\end{equation} 
gives the evolution of the wavepacket density. Inside the barrier the electron density vanishes (hence
there is no electron wavepacket) and the positron density reads
\begin{equation}
\rho_{po}(x,t)=\left\langle 0\right\vert \Psi_{po}^{\dagger}(x,t)\Psi
_{po}(x,t)\left\vert 0\right\rangle -\rho_{po}^{\zeta}(x,t), \label{pbd}%
\end{equation}
where the last term
\begin{equation}
\rho _{po}^{\zeta }(x)=\left\vert \int dpdk \zeta (p)U_{\varphi _{k}\phi _{p}}(t)\varphi _{k}(x)\right\vert ^{2}
\end{equation}
appears as a correction to the vacuum positron density due to the
incoming electron scattering on the supercritical potential \cite{SP}.

In the case of resonant Klein tunneling, the barrier
needs to be narrow relative to the wavepacket width, so this decrease will
be small, but it is nevertheless crucial in order to modulate pair production
at the right edge of the barrier. Indeed, most of the wavepacket
will typically reach the barrier at times for which the pair production rate
becomes negligible. The mechanism invoked in Ref.\ \cite{grobe2004} explaining
the Klein paradox for a supercritical step as the result of Pauli blockade
should be parsed differently for a finite barrier
since a sizeable part of the wavepacket will typically reach the barrier at
times for which the pair production rate vanishes. A mechanism consistent with our computations could
be the following: (i) The incoming electron annihilates a barrier positron \footnote{Recall that the supercritical
barrier is modeled as an external unquantized background field that creates
and annihilates particles associated with the fermionic field.}, depleting the positron density, thereby creating a dip.
(ii) The dip propagates inside the
barrier; physically, the positrons' motion is opposite to that of the dip, 
as the states of the annihilated positrons become unoccupied. (iii) Upon reaching the right edge, the dip stimulates pair
production. (iv) The created electrons at the right edge of the barrier
account for the recreated transmitted wavepacket. Note that the depletion
instability undergoes multiple reflections inside the barrier, with
successively decreasing amplitudes \cite{alkhateeb2021}. Hence when the positron density dip
reaches the left edge, pair production is stimulated and the resulting created
electronic charge contributes to the small reflected wavepacket.

\begin{figure}[t]
	\centering \includegraphics[width=8.8cm]{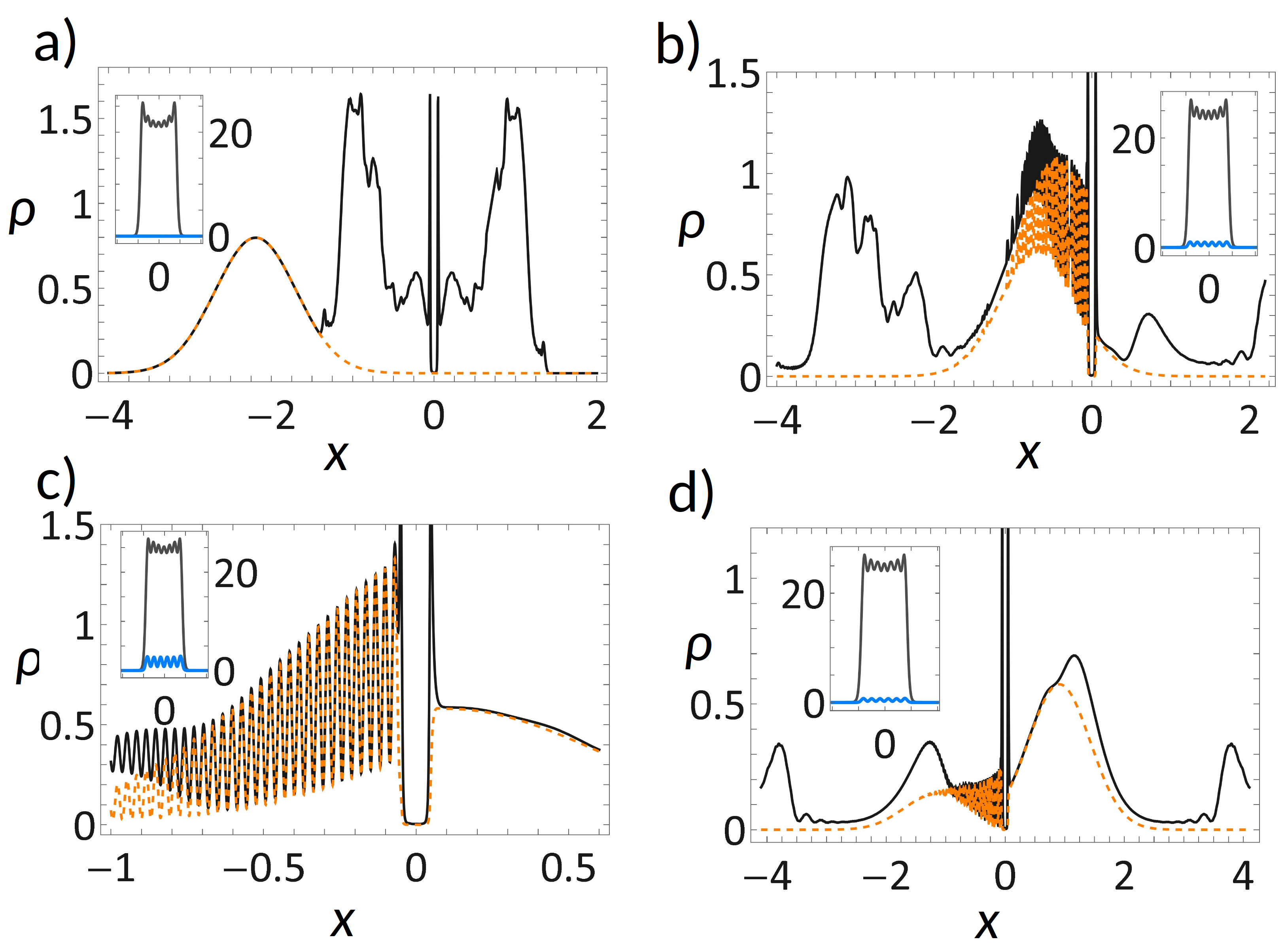} 
	\caption{Dynamics of Klein tunneling: an electron wave-packet initially ($t=0$) centered at $x=-3$ a.u. to the left of the supercritical barrier is launched with mean momentum $p_0=100$ a.u. The barrier ($V_0=3mc^2$, $\epsilon=0.3/c$) lies in the region $-0.0474<x<0.0474$ a.u., chosen so that the width $L$ obeys the resonance condition, Eq. (\ref{res-c}). (a) Snapshot of particle densities at $t=10^{-2}$ a.u., as the electron wavepacket (dotted-orange line) approaches the barrier. The solid black line gives the total electron density (due to pair creation as well as the wavepacket). The positron density lies outside the scale of the main plot and is shown in the inset (thick grey line); the thin blue line is the contribution of the modulations due to the interaction between the wavepacket and the barrier. (b) Snapshot at $t=3\times 10^{-2}$ a.u as the wavepacket reaches the barrier. (c) Zoom around the barrier region at $t=4\times 10^{-2}$ a.u., as the electron wavepacket is reformed at the right edge of the barrier; note there is no electron density inside the barrier, but the positron modulations (thin line in the inset) are of greater amplitude. (d) At $t=5\times 10^{-2}$ a.u. most of the wavepacket (dotted orange line) is ``transmitted'' to the right, while a smaller fraction is reflected}%
	\label{figkt}%
\end{figure}

\begin{figure}[t]
	\centering \includegraphics[width=6cm]{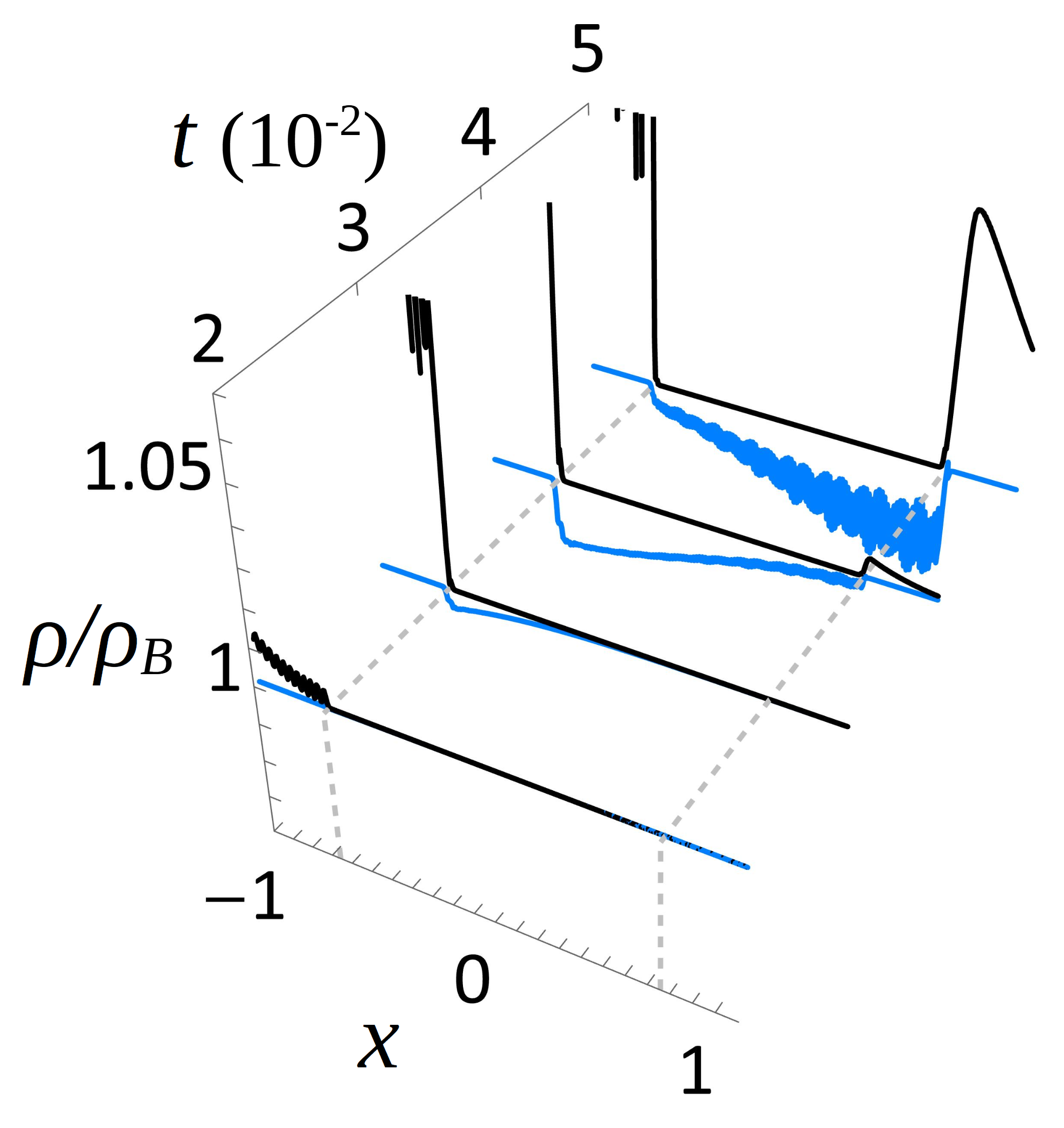} 
	\caption{The transmission mechanism detailed in the text is illustrated for an electron wavepacket colliding on a smooth rectangular barrier (region between dashed lines, $x$ in a. u.). The ratio $\rho_{el}(x,t)/\rho_{\mathrm{B}el}(x,t)$ is shown in black at different times ($\rho_{el}$ is the total electron density and $\rho_{\mathrm{B}el}$ the electron density in the absence of an incoming wavepacket). The corresponding positron number density ratio  $\rho_{pos}(x,t)/\rho_{\mathrm{B}pos}(x,t)$ is shown in gray (online blue). At $t=2\times10^{-2}$ au, the front tail of the electron wavepacket reaches the left edge of the filled barrier. This creates a dip in the positron density, visible at $t=3\times10^{-2}$ a. u. The dip propagates inside the barrier ($t=4\times10^{-2}$ au), and reaches the right edge, stimulating pair production. The additional created electron excitations (relative to vacuum polarization) to the right of the barrier are seen at $t=5\times10^{-2}$ a. u. to correspond to the front tail of the wavepacket that appears has having been transmitted.}%
	\label{fig-add}%
\end{figure}

Numerical results illustrating resonant Klein tunneling are given in Fig. \ref{figkt}.
Each panel shows snapshots at different times. By the time the initial electron
wavepacket [Fig. \ref{figkt}(a)] reaches the barrier, the pair production rate has already decreased
[Fig. \ref{figkt}(b)].\ A small part of the electron wavepacket is reflected, while the
transmitted part annihilates barrier positrons thereby modulating the positron density inside the barrier
[Fig. \ref{figkt}(c)].\ Note there is no electron inside the barrier (the electron
density is vanishingly small). The electron wavepacket is then reformed at the
right edge of the barrier [Fig.\ \ref{figkt}(d)] by pair creation due to the positron
modulation. Hence there is no tunneling in Klein tunneling, but a change in
the pair creation rate at both edges of the barrier caused by the incoming electron. Another 
example detailing the barrier dynamics for a wider and parsing each step of the proposed mechanism is shown in Fig. \ref{fig-add}.

The present QFT formalism can be applied similarly to spin-0 bosons obeying the Klein-Gordon
equation. The field operators in Eqs. (\ref{field1})-(\ref{bogu2}) now obey
commutation rules, and the basis functions $\phi_{p}$ and $\varphi_{p}$ are
2-component solutions of the free Klein-Gordon equation expressed in
Hamiltonian form \cite{greiner}. As is well-known, bosons scattering on a supercritical
potential step give rise to superradiance \cite{manogue}, whereby
the bosonic amplitude reflected from the step is larger than the incoming one \cite{grobe-analog} 
(compensated inside the step by the creation of anti-bosonic amplitude). The
same phenomenon in the case of a supercritical barrier is expected to lead to an exponential
rate of pair creation, as the anti-bosons created at one edge of the barrier
undergo multiple scatterings inside the barrier. Our numerical results, shown in Fig. 2(b) for increasing barrier widths confirm this behavior. For a fixed potential, the
momentum distribution of the ejected bosons peaks at $\left(  V_{0}^2%
-4m^{2}c^{4}\right)  ^{1/2}/2c$, so the creation rate only depends on the
time taken by the anti-bosons to travel from one edge to the other. A similar self-amplification takes place in supercritical wells \cite{grobe-su-well} and can be traced back to the divergent behavior of the scattering amplitudes when interacting with a field inhomogeneity \cite{AJP}. 

Note that in the case of bosonic Klein tunneling, the incoming boson will also undergo an amplification by multiple scattering inside the barrier; the total anti-boson ($ab$) density inside the barrier takes the form
\begin{equation}
	\rho_{ab}(x,t)=\left\langle 0\right\vert \Psi_{ab}^{\dagger}(x,t)\Psi
	_{ab}(x,t)\left\vert 0\right\rangle +\rho_{ab}^{\zeta}(x,t), \label{pbd-bos}
\end{equation}
similar to Eq. (\ref{pbd}) but with the sign of the modulation inverted relative to the fermionic case. The incoming boson now enhances pair-production when colliding at the left edge of the barrier. The modulation caused by the wavepacket, described by the term $\rho_{ab}^{\zeta}$ in Eq. (\ref{pbd-bos}), propagates inside the barrier, amplifying the charge with each collision on a barrier edge on top of the anti-bosonic charge produced by the field. 

To sum up, we have investigated Klein tunneling across a supercritical barrier employing a space-time resolved QFT approach. While this approach
does not aim at quantitative predictions for a specific future experiment, the
present nonperturbative theoretical framework provides an understanding
of the elementary processes underlying not only Klein tunneling, but also the radiating properties of the barrier. Concerning the latter, we recovered the expected asymptotic behavior for pair production by a symmetric background potential \cite{calogeracos}, but have also further unraveled the detailed time-dependent dynamics and correlations for shorter times. Our proposed mechanism describing the undamped tunneling feature when a particle impinges on a finite barrier goes beyond the results obtained previously \cite{qft-rev,grobe2004} concerning the resolution of the Klein paradox for a potential step, in that the mechanism accounts for particle transmission. Additional refinements, such as the inclusion of the Coulomb repulsion, will be necessary in order to achieve a definitive understanding of tunneling in supercriticial potentials.


\begin{thebibliography}{99}                                                                                               %


\bibitem {ruffiniR}R. Ruffini, G. Vereshchagin and S.-S. Xue, Phys. Rep. 487,
1 (2010)

\bibitem {reviewR}A. Fedotov, A. Ilderton, F. Karbstein, B. King, D. Seipt, H. Taya, and G. Torgrimsson, arXiv:2203.00019 (2022).


\bibitem {dunneR}G.\ V.\ Dunne, Eur. Phys. J. D 55, 327 (2009).

\bibitem {hu}H. Hu, Contemp. Phys. 61, 12 (2020).

\bibitem {shabaev}I.\thinspace A. Aleksandrov, G. Plunien, and V. M.
Shabaev, Phys. Rev. D 94, 065024 (2016).

\bibitem{gies} H. Gies and G. Torgrimsson, Phys. Rev. Lett. 116, 090406 (2016).

\bibitem {grobe-opt}J. Unger, S. Dong, Q. Su, and R. Grobe, Phys. Rev. A 100,
012518 (2019).



\bibitem {lv}D.\thinspace D. Su, Y.\thinspace T. Li, Q.\thinspace Z. Lv, and
J. Zhang, Phys. Rev. D 101, 054501 (2020).

\bibitem {calogeracos}N. Dombey and A. Calogeracos, Phys. Rep. 315, 41
(1999)


\bibitem {greiner}W. Greiner, B. Müller, and J. Rafelski, \emph{Quantum
Electrodynamics of Strong Fields} (Springer-Verlag, Berlin, 1985), Ch. 5 and 10.

\bibitem {alkhateeb2021}M. Alkhateeb, X. Gutierrez de la Cal, M. Pons, D.
Sokolovski and A. Matzkin, Phys. Rev. A 103, 042203 (2021).

\bibitem {nikishov}A.\ I.\ Nikishov, Phys. At. Nucl. 67 1478 (2004).

\bibitem {hansen}A. Hansen and F. Ravndal, Phys. Scripta 23, 1036 (1981).

\bibitem {gavrilov}S. P. Gavrilov and D. M. Gitman, Phys. Rev. D 93, 045002 (2016).

\bibitem {kleinert}A.\ Chervyakov and H. Kleinert, Phys. Part. Nuclei 49 374 (2018).

\bibitem {grobe2004}P. Krekora, Q. Su, and R. Grobe, Phys.\ Rev.\ Lett.\ 92
040406 (2004).

\bibitem {mckellar}M. J. Thomson and B. H. J. McKellar, Am.\ J.\ Phys. 59, 340 (1991).

\bibitem {cmo}A. Young and P.\ Kim, Nature Phys 5, 222 (2009).

\bibitem {kgs}X Jiang et al, Science 370, 1447 (2020).


\bibitem {bco}R. Gerritsma \emph{et al}, Phys. Rev. Lett. 106, 060503 (2011).


\bibitem {oo}T. Salger \emph{et al}., Phys. Rev. Lett. 107, 240401 (2011).


\bibitem {gitman-back}S. P. Gavrilov and D. M. Gitman, Phys.\ Rev.\ Lett. 101
130403 (2008).

\bibitem {qft-rev}T. Cheng, Q. Su, and R. Grobe, Contemp. Phys. 51, 315 (2010).

\bibitem {split}M. Ruf, H. Bauke and C. H. Keitel, J. Comp. Phys. 228 9092 (2009).

\bibitem {SP}See Appendix for details.

\bibitem {schweber}S. S. Schweber, \emph{An Introduction to Relativistic
Quantum Field Theory}, (Dover, Mineola, 2005).

\bibitem {add-r}P. Krekora, Q. Su, and R. Grobe, Phys. Rev. A 73, 022114 (2006).

\bibitem {oldmulti}K. Hencken, D. Trautmann, and G. Baur, Phys. Rev. A 51, 998 (1995).

\bibitem {grobe-multi}T. Cheng, Q. Su, and R. Grobe, Phys. Rev. A 80, 013410 (2009).

\bibitem {lv-bauke}Q.\thinspace Z. Lv and H. Bauke, Phys. Rev. D 96, 056017 (2017).


\bibitem {prep}M.\ Alkhateeb and A.\ Matzkin, in preparation.

\bibitem {graphene}M. I. Katsnelson, K. S. Novoselov and A. K. Geim, Nature
Phys. 2, 620 (2006).

\bibitem {keitel2009}M. Ruf, G. R. Mocken, C. Müller, K. Z. Hatsagortsyan and
C. H. Keitel, Phys.\ Rev.\ Lett. 102, 080402 (2009).


\bibitem {calogeracos2} N. Dombey, P. Kennedy and A. Calogeracos, Phys. Rev. Lett. 85, 1787 (2000).

\bibitem {manogue}C. A. Manogue, Ann. Phys. 181, 261 (1988).

\bibitem{grobe-analog}R. E. Wagner, M. R. Ware, Q. Su, and R. Grobe, Phys. Rev. A 81, 024101 (2010).

\bibitem {grobe-su-well}R. E. Wagner, M. R. Ware, Q. Su, and R. Grobe, Phys. Rev. A 81, 052104 (2010).

\bibitem {AJP}M. Alkhateeb and A. Matzkin, Am. J. Phys.  90, 297 (2022).


\end{thebibliography}
\end{document}